\documentclass{pasa}%

\title[Liberation of specific angular momentum]{Liberation of specific angular momentum through radiation and scattering in relativistic black-hole accretion discs}
\author[A.~R.~H.~Stevens]{Adam R.~H.~Stevens$^{1,2}$\thanks{astevens@swin.edu.au}\\
\affil{$^1$Centre for Astrophysics \& Supercomputing, Swinburne University of Technology, Hawthorn, VIC 3122, Australia}%
\affil{$^2$Institute of Astronomy and Kavli Institute for Cosmology, University of Cambridge, Cambridge, CB3 0HA, United Kingdom}}%
\jid{PASA}
\doi{10.1017/pas.\the\year.xxx}
\jyear{\the\year}

\usepackage[authoryear]{natbib}
\bibpunct{(}{)}{;}{a}{}{,}
\usepackage{graphicx}
\usepackage[scr=rsfso]{mathalfa}

\begin{document}%
\begin{abstract}
A key component of explaining the array of galaxies observed in the Universe is the feedback of active galactic nuclei, each powered by a massive black hole's accretion disc.  For accretion to occur, angular momentum must be lost by that which is accreted.  Electromagnetic radiation must offer some respite in this regard, the contribution for which is quantified in this paper, using solely general relativity, under the thin-disc regime.  Herein, I calculate extremised situations where photons are entirely responsible for energy removal in the disc and then extend and relate this to the standard relativistic accretion disc outlined by \citeauthor{nt}, which includes internal angular-momentum transport.  While there is potential for the contribution of angular-momentum removal from photons to be $\gtrsim$1\% out to $\sim$10$^4$ Schwarzschild radii if the disc is irradiated and maximally liberated of angular momentum through inverse Compton scattering, it is more likely of order $10^2$ Schwarzschild radii if thermal emission from the disc itself is stronger.  The effect of radiation/scattering is stronger near the horizons of fast-spinning black holes, but, ultimately, other mechanisms must drive angular-momentum liberation/transport in accretion discs.
\end{abstract}
\maketitle%
\section{INTRODUCTION}
\label{sec:intro}

It has long been widely accepted that active galactic nuclei are powered by gravitationally liberated energy from accretion discs around massive black holes \citep{lynden-bell69}.  While the study of accretion is an interesting prospect in itself, the feedback it ensues is significantly consequential for the evolution of galaxies as well \citep[e.g.][]{dimatteo05}.  To truly gauge the effect of feedback requires highly detailed cosmological hydrodynamic simulations that self-consistently track the growth of black holes and the emission from their accretion discs.  However, even the most state-of-the-art simulations presently \citep[e.g.][]{illustris,eagle} are unable to resolve accretion discs, and must use subresolution models to describe their physics \citep[e.g.][]{springel05,booth09}, in a similar vein to semi-analytic models \citep[e.g.][]{croton06,benson12}.  An analytic understanding of the functioning of accretion discs is hence key for this cause.

In the simple picture of a thin accretion disc, particles quasi-statically shrink on equatorial, circular orbits, where pressure forces are assumed negligible, until they reach the orbit of lowest energy, after which they are assumed to be captured by the black hole \citep{lynden-bell69,bardeen70}.  In doing so, those particles must be liberated of their angular momentum.  Either angular momentum is lost through the disc out to higher radii, or it is emitted vertically and removed from the disc entirely.  The latter occurs naturally through thermal emission of the disc and through scattering of photons if the disc is subject to irradiation from an external source.  It is of interest then to assess the contribution of angular-momentum liberation that photons provide in order to better understand the process of accretion itself.  This paper aims to calculate exactly this, using purely general relativistic arguments.

In Section \ref{sec:math} of this paper, relevant mathematical formulae for studying accretion discs are outlined.  General relativistic calculations are performed based on these formulae in Section \ref{sec:main}, where limiting cases for the liberation of angular momentum via photons, as well as from the standard \citet{nt} disc, are considered.  Concluding remarks are provided in Section \ref{sec:conclusion}.

\section{MATHEMATICAL FORMALISMS AND BACKGROUND}
\label{sec:math}

The unique metric for spacetime around a (non-charged) rotating source mass (e.g.~a black hole) was first discovered by \citet{kerr63}, usually now written in Boyer-Linquist coordinates \citep{boyer67}, for which the invariant interval is
\begin{subequations}
\label{eq:interval}
\begin{multline}
-c^2 \mathrm{d} \tau^2 = -\left( 1 - \frac{r_s r}{\rho} \right) c^2 \mathrm{d} t^2 + \frac{\rho}{\Delta} \mathrm{d} r^2 + \rho\ \mathrm{d} \theta^2 + \\ \left( r^2 + a^2 + \frac{r_s r a^2 \sin^2(\theta)}{\rho} \right) \sin^2(\theta)\  \mathrm{d} \phi^2 \\ - \frac{2 r_s r a^2 \sin^2(\theta)}{\rho}\ c\ \mathrm{d} t\ \mathrm{d}\phi\ ;
\end{multline}
\begin{equation}
\rho \equiv r^2 + a^2 \cos^2(\theta)\ ,\ \ \ \Delta \equiv r^2 - r_s r + a^2\ ,
\end{equation}
\end{subequations}
where $c$ is the speed of light, $r_s \equiv 2GM/c^2$ is the Schwarzschild radius, and $a \equiv J/Mc$ is the spin parameter (specific angular momentum) of the source of mass $M$ and angular momentum $J$.  The metric is stationary, axisymmetric, and, as \citet{carter68} showed \citep[but see also][\S33.5]{gravitation}, exhibits four constants of motion.\footnote{In fact, those authors showed this for the more general, charge-inclusive Kerr-Newman metric \citep{newman65}.}  Two of these constants are the azimuthal and time components of covariant four-momentum.  Taking the limit $r \rightarrow \infty$, one finds these to be relativistic analogues of energy and azimuthal angular momentum, usually referred to as the energy and angular momentum ``at infinity''.  Often the ``at infinity'' is dropped for brevity, and the usual symbols for these quantities are used, i.e.
\begin{equation}
\label{eq:ELdef}
E \equiv -p_t\ ,\ \ \ L_z \equiv p_{\phi} \ .
\end{equation}
When discussing the emission or transport of energy or angular momentum in accretion discs (or Kerr geometry in general), these are the quantities that are meant.

Throughout the rest of this paper, most quantities will be expressed in a dimensionless form, represented by a bar placed on the quantity of interest.  For quantities with dimensions of distance, this means normalising to half the Schwarzschild radius, e.g.~$\bar{r} \equiv 2r/r_s$, $\bar{a} \equiv 2a/r_s$, in line with literature convention.  Equation \ref{eq:EL} covers quantities with other dimensions.

By analysing equations of motion for particles in a Kerr spacetime, \citet{bardeen72} obtained expressions for the energy and specific angular momentum for circular (i.e.~$p^r=0$), equatorial (i.e.~$\theta=\pi/2$ and $p^{\theta}=0$), Keplerian (i.e.~gravity is entirely centrifugally balanced) orbits:
\begin{subequations}
\label{eq:EL}
\begin{equation}
\bar{E} \equiv \frac{E}{mc^2} = \frac{\bar{r}^{3/2} - 2\bar{r}^{1/2} \pm \bar{a}}{\bar{r}^{3/4} \left( \bar{r}^{3/2} - 3 \bar{r}^{1/2} \pm 2\bar{a} \right)^{1/2}}\ ,
\end{equation}
\begin{equation}
\bar{L}_z \equiv \frac{L_z}{mcr_s} = \frac{\pm\bar{r}^2 - 2\bar{a}\bar{r}^{1/2} \pm \bar{a}^2}{2\bar{r}^{3/4} \left( \bar{r}^{3/2} - 3\bar{r}^{1/2} \pm 2\bar{a} \right)^{1/2}}\ ,
\end{equation}
\end{subequations}
where upper signs are for prograde orbits and lower signs retrograde.  Note that $m$ represents the \emph{rest} mass of a particle (at infinity), \emph{not} its \emph{inertial} mass.  By checking the derivatives of these quantities,
\begin{subequations}
\label{eq:dEL}
\begin{equation}
\frac{\mathrm{d} \bar{E}}{\mathrm{d} \bar{r}} = \frac{8\bar{a} \bar{r}^{1/2} - 3\bar{a}^2 + \bar{r}(\bar{r}-6)}{W}\ ,
\end{equation}
\begin{multline}
\frac{\mathrm{d} \bar{L}_z}{\mathrm{d} \bar{r}} = W^{-1} \Bigg\{ \bar{a}^2 \bar{r}^{1/2} \left(4-\frac{3}{2}\bar{r}\right) + \bar{r}^{5/2}\left(\frac{1}{2}\bar{r}-3\right) \\ - 3\left[\frac{1}{2}\bar{a}^3 - \bar{a}\bar{r}\left(\frac{3}{2}\bar{r}-1\right)\right] \Bigg\}\ ;
\end{multline}
\begin{equation}
W \equiv 2 \bar{r}^{7/4} \left[2\bar{a} + \bar{r}^{1/2} (\bar{r}-3) \right]^{3/2}\ ,
\end{equation}
\end{subequations}
one finds these two relations share a common minimum, referred to in the literature as the innermost stable circular orbit (ISCO), where $\bar{r}_{\rm ISCO}$ is given by equation 2.21 of \citet*{bardeen72}.\footnote{In their notation, $r_{\rm ms}/M$ and $a/M$ are equivalent to $\bar{r}_{\rm ISCO}$ and $\bar{a}$ here, respectively.}

Under the picture where particles transit between infinitesimally adjacent orbits in the process of accretion (until they reach the ISCO), the above equations provide the starting point for calculating how much specific angular momentum can be lost from photon emission and/or scattering.  Hereafter, the use of $E$ and $L_z$ (with or without bars, but without further subscripts) refers to the orbiting states for which these equations (\ref{eq:EL}--\ref{eq:dEL}) apply.

\section{RADIATING AND SCATTERING AWAY ANGULAR MOMENTUM}
\label{sec:main}
In each of the following subsections, the relative specific-angular-momentum loss to photons in an accretion disc, $-\mathrm{d} \bar{L}_{z, \gamma} / \mathrm{d} \bar{L}_z$ (where subscript $\gamma$ is for photons), as a function of radius, is calculated for a different idealised situation, where each builds on the last.  Each result is plotted in Fig.~\ref{fig:dL} for $r \ge r_{\mathrm{ISCO}}$, allowing for comparisons between each individual case.  The models considered here are all of thin, relativistic accretion discs, whereby the mathematics of Section \ref{sec:math} is applicable.

\subsection{Pure, relatively isotropic emission}
\label{ssec:isotropic}
Before considering consequences for other forms of energy transport, the energy carried away by a photon supplied by a particle moving to an adjacent lower-energy circular orbit should, at most, be the energy difference between the orbits.  This can be written in terms of differentials as
\begin{equation}
\label{eq:EE}
\mathrm{d} E_{\gamma} = -\mathrm{d} E\ .
\end{equation}
If this energy is lost primarily through radiation, one expects each photon to be emitted with statistical isotropy from the frame of the emitting particle.  For a non-rotating disc, this would make the average direction of emission from each face of the disc vertical (i.e.~initially completely in the $\pm\theta$-direction).  For a rotating disc perceived from an external frame (i.e.~one static with the Boyer-Lindquist coordinates), this can then be modelled by stating that photons are emitted in the $\phi$--$\theta$ ``plane'' with a three-velocity in the $\phi$-direction equivalent to that of the disc, naturally a function of radius.  This (angular) velocity is found as $\mathcal{V}^{\phi} = p^{\phi} / p^t$.  Recognising $p^t = -g^{tt}E + g^{t\phi}L_z$ and $p^{\phi} = -g^{t\phi}E + g^{\phi\phi}L_z$ (cf.~Equations \ref{eq:interval} and \ref{eq:ELdef}), obtaining the contravariant metric components by taking the matrix inverse of Equation \ref{eq:interval}, and expanding and simplifying with Equation \ref{eq:EL}, one concludes consistently with \citet*{bardeen72} that
\begin{equation}
\label{eq:omega}
\frac{p^{\phi}}{p^t} = \frac{\pm(2r_s)^{1/2}c}{2r^{3/2} \pm (2r_s)^{1/2} a}\ .
\end{equation}

\begin{figure}[t]
\centering
\includegraphics[width=0.48\textwidth]{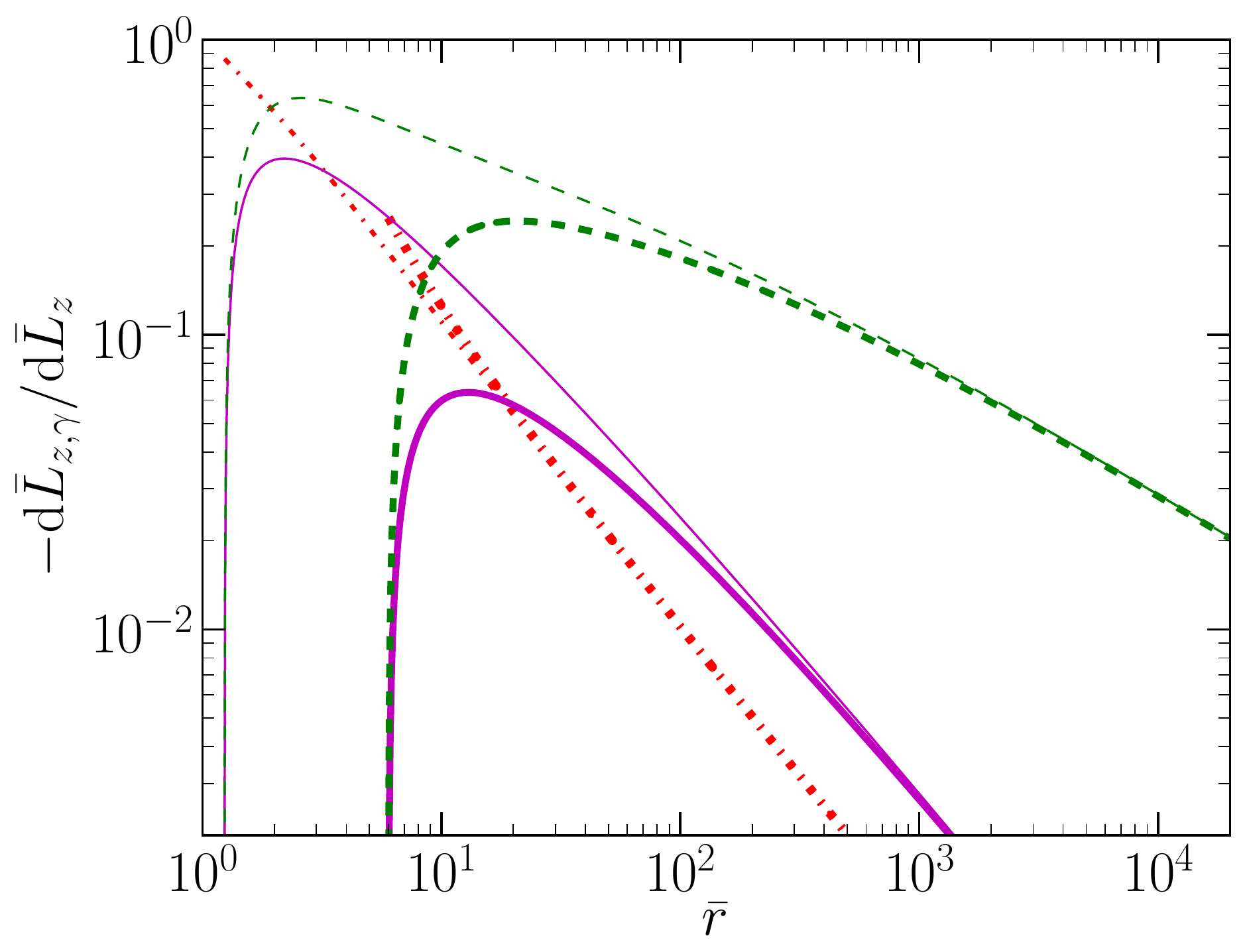}
\caption{(Specific) angular momentum lost to photons relative to the (specific-)angular-momentum gap between adjacent, equatorial, circular orbits in a relativistic accretion disc as a function of radius.  Thick curves apply for an accretion disc around a non-spinning black hole, while thin curves are for a hole spinning with $\bar{a} = 0.998$ \citep[the maximum of][]{thorne74}.  The dot-dashed curves assume photons are emitted with energy equal to the difference of the orbits and are angled to the accretion disc plane such that the $\phi$-velocity of the photons matches that of the disc itself (see Section \ref{ssec:isotropic}).  The solid curves follow the solution of \citet{nt}, which include effects of internal torques (outlined further in Section \ref{ssec:viscosity}).  The dashed curves also account for internal torques, and show the upper limit of scattering, where the momentum imparted on photons is parallel to the $\phi$-direction (Section \ref{ssec:scattering}).}
\label{fig:dL}
\end{figure}

Let us write the (contravariant) four-momentum components of an emitted photon as $\mathrm{d} p^{\mu}$.  One can then simultaneously solve
\begin{equation}
\label{eq:dEg}
g_{tt} \mathrm{d} p^t + g_{t\phi} \mathrm{d} p^{\phi} = -\mathrm{d} E_{\gamma}
\end{equation}
and Equation \ref{eq:omega} (for the latter, the left-hand side now reads $\mathrm{d} p^{\phi}/\mathrm{d} p^t$) to obtain explicit functions for $\mathrm{d} p^t$ and $\mathrm{d} p^{\phi}$.  Further calculating $\mathrm{d} p_{\phi} = \mathrm{d} L_{z, \gamma}$, one obtains
\begin{equation}
\label{eq:dLE_iso}
\frac{\mathrm{d} \bar{L}_{z, \gamma}}{\mathrm{d} \bar{E}_{\gamma}} = \frac{\bar{L}_z}{\bar{E}} = \frac{\pm \bar{r}^2 - 2\bar{a}\bar{r}^{1/2} \pm \bar{a}^2 }{2\left(\bar{r}^{3/2} - 2\bar{r}^{1/2} \pm \bar{a}\right)}\ ,
\end{equation}
consistent with the report of \citet[][\S 3.10]{lynden-bell86}. Now through Equation \ref{eq:EE}, combined with use of Equation \ref{eq:dEL}, one finds an explicit form of $-\mathrm{d} \bar{L}_{z, \gamma} / \mathrm{d} \bar{L}_z$, as presented by the dot-dashed lines in Fig.~\ref{fig:dL}.  For non-spinning black holes, the analytic relation simplifies to
\begin{equation}
\frac{\mathrm{d} \bar{L}_{z, \gamma}}{\mathrm{d} \bar{L}_z} (\bar{a}=0) = \frac{1}{\bar{r}-2}\ ,
\end{equation}
which further reduces to the Newtonian case presented by \citet{johnson11} for $r \gg r_s$.

Under this picture, one finds that specific-angular-momentum removal by photons is important beyond the percent level out to $\sim$$50 r_s$.  For non-spinning holes, as particles approach the ISCO, the radiative efficiency of angular momentum approaches 25\%, while for maximally spinning black holes, the efficiency approaches 87\%.  Already from this analysis, it is clear, and perhaps unsurprising, that radiation is insufficient by itself to liberate an accretion disc of its necessary specific angular momentum.

\subsection{Relatively isotropic emission with internal angular-momentum transport}
\label{ssec:viscosity}
The previous subsection considered a limiting role of photons without any additional form of angular-momentum transport.  Because photons are unable to remove all the necessary angular momentum if they remove all the necessary energy, some other mechanism must remove angular momentum, which consequently must alter the energy liberated by photons too.

The standard model of thin accretion discs proposed by \citet{ss73}, and extended to be relativistic by \citet{nt}, considers angular momentum to be transported radially through internal torques \citep[generated by magnetically induced ``viscosity'' -- see][]{lynden-bell69,ss73,lin13}, in addition to removal from photons, in an accretion disc whose structure is completely stable (i.e.~not a function of time).  If $\mathrm{d} \bar{L}_{z, \mathrm{int}} / \mathrm{d} \bar{r}$ represents the radial \emph{net removal} of specific angular momentum from internal torques during a particle transition to an infinitesimally adjacent orbit, then it must hold that
\begin{equation}
\label{eq:LLL}
\frac{\mathrm{d} \bar{L}_z}{\mathrm{d} \bar{r}} + \frac{\mathrm{d} \bar{L}_{z, \mathrm{int}}}{\mathrm{d} \bar{r}}  + \frac{\mathrm{d} \bar{L}_{z, \gamma}}{\mathrm{d} \bar{r}}  = 0\ .
\end{equation}
With some small rearranging, the solution derived purely from the continuity equations of rest mass, angular momentum, and energy by \citet{nt} provides
\begin{subequations}
\label{eq:nt}
\begin{equation}
\frac{\mathrm{d} \bar{L}_{z, \mathrm{int}}}{\mathrm{d} \bar{r}}  = - \frac{1}{2} \frac{\mathrm{d}}{\mathrm{d} \bar{r}} \left( \sqrt{\bar{r}} \mathscr{Q} \right)\ ,
\end{equation}
\begin{equation}
\frac{\mathrm{d} \bar{L}_{z, \gamma}}{\mathrm{d} \bar{r}}  = - \frac{3}{2}  \frac{\bar{L}_z}{\bar{r}^2} \frac{\mathscr{Q}}{\mathscr{B}\sqrt{\mathscr{C}}}\ ,
\end{equation}
\end{subequations}
where $\mathscr{Q}$, $\mathscr{B}$, and $\mathscr{C}$ are dimensionless quantities that approach unity for increasing $r$: the reader is referred to equation 5.4.1 of \citet{nt} for their formal definitions.\footnote{\citet{pt74} note a sign error for equation 5.4.1h.  Those authors also provide an analytically integrated expression for $\mathscr{Q}$, but using that provides different results and does not satisfy Equation \ref{eq:LLL}, whereas numerically integrating the equations of \citet{nt} does.  For an alternate parametrisation, see \citet{riffert95}.}

As for the previous subsection, $-\mathrm{d} \bar{L}_{z, \gamma} / \mathrm{d} \bar{L}_z$ is presented for the \citet{nt} model in Fig.~\ref{fig:dL} as solid lines.  For non-spinning black holes, the liberation of angular momentum from photons at low radii is noticeably less effective than the previous limiting case.  Faster spinning holes reach a peak efficiency of radiative angular-momentum removal of nearly 40\%.

\begin{figure}[t]
\centering
\includegraphics[width=0.48\textwidth]{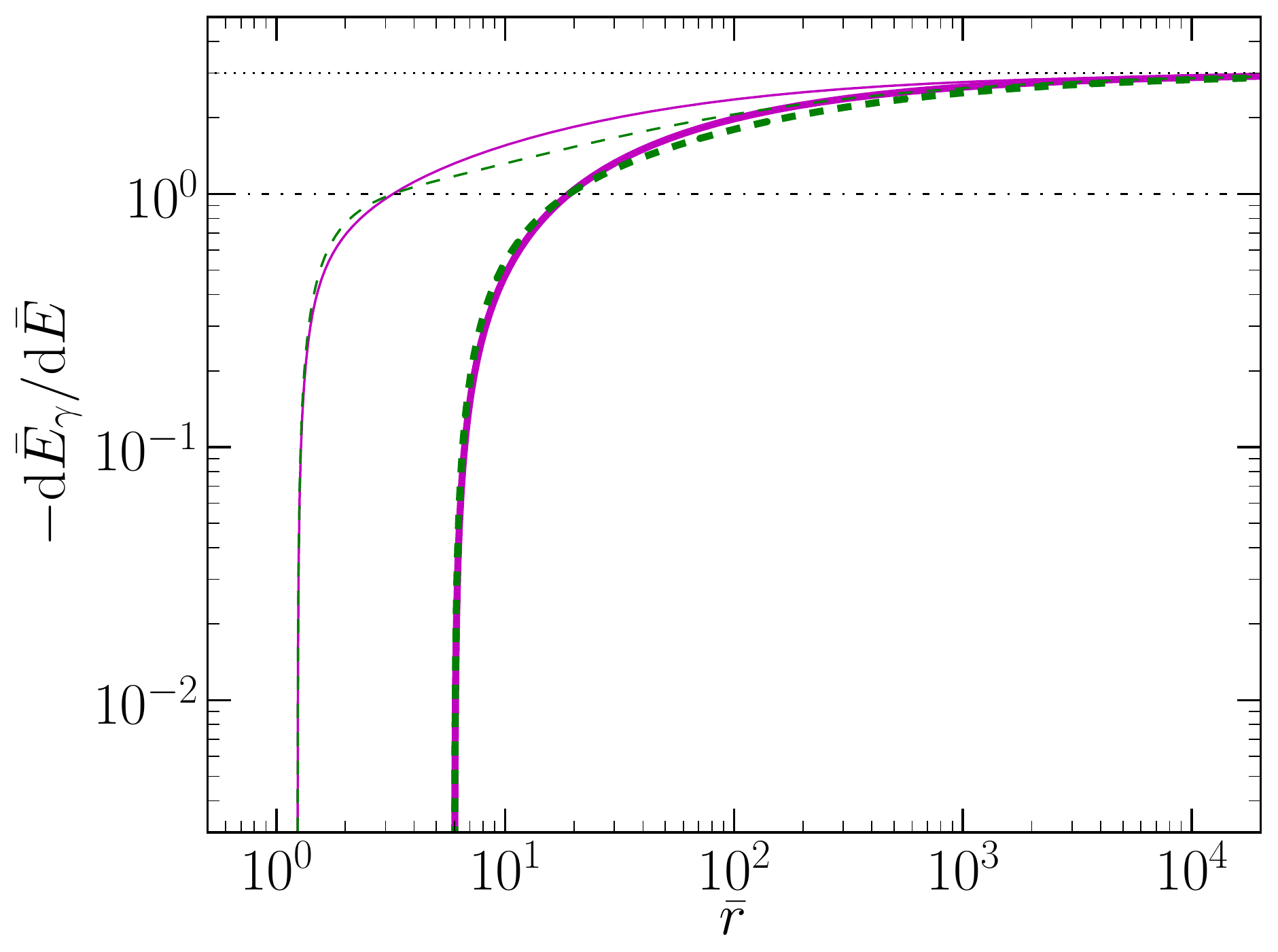}
\caption{Relative energy scattered away or emitted through photons between infinitesimally adjacent orbits for a \citet{nt} accretion disc (solid curves, Section \ref{ssec:viscosity}) and a maximally scattering-dominant disc with internal angular-momentum transport (dashed curves, Section \ref{ssec:scattering}) around non-spinning (thick curves) and maximally spinning (thin curves) black holes.  Where the curves pass above a value of 1 (dot-dashed line), extra energy is radiated away, transferred internally within the disc.  The dotted line indicates an asymptote; at large radii, a massive particle in the disc emits a photon with thrice the necessary energy to reach its adjacent lower orbit to account for energy supplied by internal transport.}
\label{fig:Eph}
\end{figure}

Na\"{i}vely, one may have expected the solid lines in Fig.~\ref{fig:dL} to lie underneath the dot-dashed lines (i.e.~the result of Section \ref{ssec:isotropic}), because a new mode of angular-momentum transport has been introduced since Section \ref{ssec:isotropic}.  However, as noted by \citet{ss73}, internal angular-momentum transport provides a net energy \emph{source} for $r \gg r_{\rm ISCO}$, meaning more angular momentum must be liberated by photons, specifically by a factor of 3.  This is reconcilable by considering an energy outflow balance equation (i.e.~an energy version of Equation \ref{eq:LLL}),
\begin{equation}
\label{eq:EEE}
\frac{\mathrm{d} \bar{E}}{\mathrm{d} \bar{r}} + \frac{\mathrm{d} \bar{E}_{\mathrm{int}}}{\mathrm{d} \bar{r}}  + \frac{\mathrm{d} \bar{E}_{\gamma}}{\mathrm{d} \bar{r}}  = 0\ .
\end{equation}
Using a combination of Equations \ref{eq:dEL}a, \ref{eq:dLE_iso}, and \ref{eq:nt}b, one can determine $-\mathrm{d} \bar{E}_{\gamma} / \mathrm{d} \bar{E}$ to show that indeed this asymptotes to a value of 3.  This is plotted in Fig.~\ref{fig:Eph} (solid curves). When interpreting Fig.~\ref{fig:Eph}, one should appreciate that the \emph{absolute} energy gap between adjacent orbits tends to zero as $\bar{r} \rightarrow \infty$, and that thin discs have higher densities toward their centres.

\subsection{Scattering with internal angular-momentum transport}
\label{ssec:scattering}
If an accretion disc is irradiated by an external source, incoming photons can be absorbed or scattered, causing particles in the disc to lose angular momentum, \`{a} la the Poynting-Robertson effect \citep{poynting04,robertson37,burns79}.  In the absorption case, where the motive absorbing particles re-emit the radiation, the analysis of Section \ref{ssec:viscosity} remains sound.  However, it could also transpire that charged particles in the disc anisotropically transfer their energy to the incoming photons via inverse Compton scattering.  

An external irradiative source is observationally motivated by X-ray reflection spectra generated by an accretion disc's corona, which provide a means for measuring black holes' spins \citep[for a review, see][]{reynolds14}.  At very high redshift, the cosmic background radiation could also provide an irradiative source \citep[e.g.][]{fukue94,mineshige98}, although far more modest in temperature.  For fast-spinning holes, a notable portion of emitted radiation from accretion discs is expected to fall back on the discs as well \citep{cunningham76}.  So long as the Thomson regime is applicable in the scattering particle's reference frame, the scattered photon can have its energy significantly multiplied and be beamed in the direction of the scatterer's motion as perceived by an external observer \citep[see][\S 7.1]{rybicki79}.  While the precise direction of the photon's change of momentum would depend on its initial energy relative to the scattering particle, one can consider the extreme upper limit where this is the $\phi$-direction, as shown immediately below.

Let us now write the four-momentum components of that imparted on the scattered photon as $\mathrm{d} p^{\mu}$ [it is very important to note that this is the \emph{change} in photon's momentum from scattering; when considering radiation, this was the momentum of the (average) photon itself, as it did not exist prior (i.e.~it had no initial momentum)].  Because photons have no rest mass, it must hold that $g_{\mu \nu} \mathrm{d} p^{\mu} \mathrm{d} p^{\nu} = 0$, where the $g_{\mu \nu}$ terms are the metric components obtainable from Equation \ref{eq:interval}, hence
\begin{equation}
\label{eq:sim}
g_{tt}(\mathrm{d} p^t)^2 + 2g_{t\phi} \mathrm{d} p^t \mathrm{d} p^{\phi} + g_{\phi\phi}(\mathrm{d} p^{\phi})^2 = 0\ .
\end{equation}
Simultaneously solving Equations \ref{eq:dEg} and \ref{eq:sim}, recognising $\mathrm{d} L_{z, \gamma} = \mathrm{d} p_{\phi} = g_{t \phi} \mathrm{d} p^t + g_{\phi\phi} \mathrm{d} p^{\phi}$ in this maximal limit (cf.~Equation \ref{eq:ELdef}), and using Equation \ref{eq:EE}, one obtains
\begin{equation}
\label{eq:dLE_phi}
\frac{\mathrm{d} \bar{L}_{z, \gamma}}{\mathrm{d} \bar{E}_{\gamma}} = -\frac{\mathrm{d} \bar{L}_{z, \gamma}}{\mathrm{d} \bar{E}} = \frac{\pm \bar{r} \sqrt{\bar{a}^2 + \bar{r}^2 - 2\bar{r}} - 2\bar{a}}{2\bar{r} - 4}\ .
\end{equation}

One can consider taking the same energy lost to photons from Section \ref{ssec:viscosity} but instead using it to kick (scatter) photons in the $\phi$-direction.  The relative increase in specific-angular-momentum loss can then be found by taking the ratio of Equation \ref{eq:dLE_phi} to Equation \ref{eq:dLE_iso}.  However, by increasing $\mathrm{d} \bar{L}_{z, \gamma}/ \mathrm{d} \bar{r}$, it must be true that $\mathrm{d} \bar{L}_{z, \mathrm{int}} / \mathrm{d} \bar{r}$ decreases, in order to satisfy conservation of angular momentum (Equation \ref{eq:LLL}).  If $\mathrm{d} \bar{L}_{z, \mathrm{int}} / \mathrm{d} \bar{r}$ decreases, then so must $\mathrm{d} \bar{E}_{\mathrm{int}} / \mathrm{d} \bar{r}$ by an amount found by taking the ratio of Equation \ref{eq:LLL} to Equation \ref{eq:EEE} after making the internal-torque terms the arguments for each:
\begin{equation}
\frac{\mathrm{d} \bar{L}_{z, \mathrm{int}}}{\mathrm{d} \bar{E}_{\mathrm{int}}} = \frac{\mathrm{d} \bar{L}_z / \mathrm{d} \bar{r} + (\mathrm{d} \bar{L}_{z, \gamma} / \mathrm{d} \bar{r})_{\ref{ssec:viscosity}}}{\mathrm{d} \bar{E} / \mathrm{d} \bar{r} + (\mathrm{d} \bar{E}_{\gamma} / \mathrm{d} \bar{r})_{\ref{ssec:viscosity}}}\ ,
\end{equation}
where subscript \ref{ssec:viscosity} implies the quantities as determined from Section \ref{ssec:viscosity}.  Consequently $\mathrm{d} \bar{E}_{\gamma} / \mathrm{d} \bar{r}$ must increase from energy conservation (Equation \ref{eq:EEE}), and therefore $\mathrm{d} \bar{L}_{z, \gamma}/ \mathrm{d} \bar{r}$ must be higher than initially calculated.  One can iteratively work through these calculations until finding a converged result.\footnote{One can update the terms with subscript \ref{ssec:viscosity} in the iterations, but the end result is the same.}

Using the above method, one can calculate the maximum $-\mathrm{d} \bar{L}_{z, \gamma} / \mathrm{d} \bar{L}_z$ for photon scattering with the effects of internal transport included.  This is shown by the dashed lines in Fig.~\ref{fig:dL}.  Consistent with the above results, under this maximal regime, accretion discs can remain efficient above the percent level for $-\mathrm{d} \bar{L}_{z, \gamma} / \mathrm{d} \bar{L}_z$ beyond $10^4 r_s$.

As was the case for the standard \citet{nt} disc, $-\mathrm{d} \bar{E}_{\gamma} / \mathrm{d} \bar{E}$ asymptotically approaches a value of 3 for increasing radii, but does so more slowly.  It also exceeds a value of 1 at the same radius, as displayed by the dashed curves in Fig.~\ref{fig:Eph}.

It should again be stressed that the calculations in this subsection are an upper limit. In truth, one should expect the relevant curve on Fig.~\ref{fig:dL} to lie between the solid and dashed ones presented, with a bias towards the former.

\section{CONCLUSION}
\label{sec:conclusion}
As material accretes onto massive black holes, there must be a process by which specific angular momentum is removed from the system.  As accretion discs are known to be bright sources of radiation, the emission of photons provides one channel for this angular-momentum loss.  If a disc is irradiated, inverse Compton scattering provides another channel.  This paper has provided calculations of the contribution of angular-momentum liberation through photons in thin, relativistic accretion discs. In addtion to situations where transport by internal torques was included, such that the usual conservation laws of physics were satisfied, these calculations included limiting situations where photon emission was responsible for all the necessary energy removal and where photon scattering was angled to remove maximal angular momentum.

On scales up to the order of a hundred Schwarzschild radii, photons remove a small ($>1$) percentage of angular momentum in accretion discs that emit in an expected fashion, but beyond this contribute negligibly.  At the absolute most, discs subject to strong irradiation are potentially capable of scattering away angular momentum as efficiently out two orders of magntiude farther from the black hole.  The contribution in both cases becomes stronger near the horizon of a fast rotating black hole, especially in the latter case ($\lesssim$60\%, cf.~thin dashed line, Fig.~\ref{fig:dL}).  By and large though, angular momentum is transported far more efficiently through the disc internally, rather than liberated from the disc electromagnetically.

\section*{ACKNOWLEDGEMENTS}
Thanks must go to Donald Lynden-Bell, Mitch Begelman, and especially the anonymous referee for their constructive thoughts on this manuscript, which inevitably lead to improvement of its content.  Thanks also to Darren Croton for his continued academic support during the publishing of this paper.


\begin{thebibliography}{}
\bibitem[Bardeen(1970)]{bardeen70}Bardeen, J.~M. 1970, Natur, 226, 64
\bibitem[Bardeen, Press \& Teukolsky(1972)Bardeen et al.]{bardeen72}Bardeen, J.~M., Press, W.~H., \& Teukolsky, S.~A. 1972, ApJ, 178, 347
\bibitem[Benson(2012)]{benson12}Benson, A.~J. 2012, NewA, 17, 175
\bibitem[Booth \& Schaye(2009)]{booth09}Booth, C.~M., \& Schaye, J. 2009, MNRAS, 398, 53
\bibitem[Boyer \& Lindquist(1967)]{boyer67}Boyer, R.~H., \& Lindquist, R.~W. 1967, JMP, 8, 265
\bibitem[Burns, Lamy \& Soter(1979)Burns et al.]{burns79}Burns, J.~A., Lamy, P.~L., \& Soter, S. 1979, Icar, 40, 1
\bibitem[Carter(1968)]{carter68}Carter, B. 1968, PhRv, 174, 1559
\bibitem[Croton et al.(2006)]{croton06}Croton, D.~J., Springel, V., White, S.~D.~M., et al. 2006, MNRAS, 365, 11
\bibitem[Cunningham(1976)]{cunningham76}Cunningham, C. 1976, ApJ, 208, 534
\bibitem[Di Matteo, Springel \& Hernquist(2005)Di Matteo et al.]{dimatteo05}Di Matteo, T., Springel, V., \& Hernquist, L. 2005, Natur, 433, 604
\bibitem[Fukue \& Umemura(1994)]{fukue94}Fukue, J., \& Umemura, M. 1994, PASJ, 46, 87
\bibitem[Johnson(2011)]{johnson11}Johnson, J.~L. 2011, AN, 332, 841
\bibitem[Kerr(1963)]{kerr63}Kerr, R.~P. 1963, PhRvL, 11, 237
\bibitem[Lin, Liu \& Xiaoqing(2013)]{lin13}Lin, F., Liu, S., \& Xiaoqing, L. 2013, NewA, 21, 40
\bibitem[Lynden-Bell(1969)]{lynden-bell69}Lynden-Bell, D. 1969, Natur, 223, 690
\bibitem[Lynden-Bell(1986)]{lynden-bell86}Lynden-Bell, D. 1986, in \emph{Gravitation in Astrophysics, Carg\`{e}se 1986} eds. B.~Carter \& J.~B.~Hartle (New York: Plenum Press)
\bibitem[Mineshige, Tsuribe \& Umemura(1998)Mineshige et al.]{mineshige98}Mineshige, S., Tsuribe, T., \& Umemura, M. 1998, PASJ, 50, 233
\bibitem[Misner, Thorne \& Wheeler(1973)Misner et al.]{gravitation}Misner, C.~W., Thorne, K.~S., \& Wheeler, J.~A. 1973, \textit{Gravitation} (New York: W.~H. Freeman and Company)
\bibitem[Newman et al.(1965)]{newman65}Newman, E.~T., Couch, E., Chinnapared, K., et al. 1965, JMP, 6, 918
\bibitem[Novikov \& Thorne(1973)]{nt}Novikov, I.~D., \& Thorne, K.~S. 1973, in \emph{Black Holes} eds. C.~DeWitt \& B.~S.~DeWitt (Paris: Gordon \& Breach)
\bibitem[Page \& Thorne(1974)]{pt74}Page, D.~N., \& Thorne, K.~S. 1974, ApJ, 191, 499
\bibitem[Poynting(1904)]{poynting04}Poynting, J.~H. 1904, RSPTA, 202, 525
\bibitem[Reynolds(2014)]{reynolds14}Reynolds, C.~S. 2014, SSRv, 183, 277
\bibitem[Riffert \& Herold(1995)]{riffert95}Riffert, H., \& Herold, H. 1995, ApJ, 450, 508
\bibitem[Robertson(1937)]{robertson37}Robertson, H.~P. 1937, MNRAS, 97, 423
\bibitem[Rybicki \& Lightman(1979)]{rybicki79}Rybicki, G.~B., \& Lightman, A.~P. 1979, \textit{Radiative Processes in Astrophysics} (Weinheim: WILEY-VCH)
\bibitem[Schaye et al.(2015)]{eagle}Schaye, J., Crain, R.~A., Bower, R.~G., et al. 2015, MNRAS, 446, 521
\bibitem[Shakura \& Sunyaev(1973)]{ss73}Shakura, N.~I., \& Sunyaev, R.~A. 1973, A\&A, 24, 337
\bibitem[Springel, Di Matteo \& Hernquist(2005)Springel et al.]{springel05}Springel, V., Di Matteo, T., \& Hernquist, L. 2005, MNRAS, 361, 776
\bibitem[Thorne(1974)]{thorne74}Thorne, K.~S. 1974, ApJ, 191, 507
\bibitem[Vogelsberger et al.(2014)]{illustris}Vogelsberger, M., Genel, S., Springel, V., et al. 2014, Natur, 509, 177
\end{thebibliography}
\end{document}